\documentclass[a4paper, amsfonts, amssymb, amsmath, reprint, showkeys, nofootinbib, pra, aps]{revtex4-2}
\usepackage[english]{babel}
\usepackage[utf8]{inputenc}
\usepackage[colorinlistoftodos, color=green!40, prependcaption]{todonotes}
\usepackage{amsthm}
\usepackage{mathtools}
\usepackage{multirow}
\usepackage{physics}
\usepackage{xcolor}
\usepackage{booktabs}
\usepackage{graphicx}
\usepackage{algorithm}
\usepackage{algorithmic}
\usepackage[a4paper,margin=2cm]{geometry}
\usepackage{adjustbox}
\usepackage{placeins}
\usepackage{listings} % Use listings instead
\usepackage{xcolor}
\definecolor{LightGray}{gray}{0.95}
\usepackage[T1]{fontenc}
\usepackage{lipsum}
\usepackage{csquotes}
\begin{document}
\title{Coined Quantum Walks on Complex Networks for Quantum Computers}

\author{Rei Sato}
    \email[Correspondence email address: ]{rei@classiq.io}% Your name
    \affiliation{Classiq Technologies G.K.,
  1-5-1 Marunouchi, Chiyoda-ku, Tokyo, 100-6509, Japan}
\date{\today}

\begin{abstract}
We propose a quantum circuit design for implementing coined quantum walks on complex networks. In complex networks, the coin and shift operators depend on the varying degrees of the nodes, which makes circuit construction more challenging than for regular networks.  To address this issue, we use a dual-register encoding to enable a simplified shift operator and reduces the resource overhead.  We implement the circuit using Qmod, a high-level quantum programming language, and evaluated the performance through numerical simulations on Erd\H{o}s–R\'enyi, Watts–Strogatz, and Barab\'asi–Albert models. The results show that the circuit depth scales as approximately $N^{1.9}$ regardless of the network topology. Furthermore, we execute the proposed circuits on the \texttt{ibm\_torino} superconducting quantum processor for Watts–Strogatz models with $N=4$ and $N=8$.  The experiments show that hardware-aware optimization slightly improved the variation distance and Hellinger fidelity for the larger network, whereas connectivity constraints imposed overhead for the smaller one.  These results indicate that while current NISQ devices are limited to small-scale validations, the polynomial scaling of our framework makes it suitable for larger-scale implementations in the fault-tolerant quantum computing era.
\end{abstract}

\keywords{quantum walk, complex network}

\maketitle

\section{Introduction}
\label{sec:introduction}
Quantum walks have emerged as one of the promising quantum gate-based algorithms for a wide range of potential use cases, including combinatorial optimization~\cite{marsh2020combinatorial}, financial modeling~\cite{orrell2021quantum}, and protein folding~\cite{casares2022qfold}.  The quantum walks are known as the quantum counterparts of classical random walks and are broadly categorized into two models: continuous-time quantum walks (CTQWs)~\cite{PhysRevA.70.022314} and discrete-time quantum walks (DTQWs)~\cite{PhysRevA.48.1687}.  While CTQWs describe network dynamics through Hamiltonian evolution, DTQWs evolve via different unitary operators~\cite{Szegedy_walk, 10.5555/1070432.1070590}.  In particular, the interplay between quantum walks and complex networks provides a variety of applications, including quantum search~\cite{PhysRevA.70.022314,10.5555/1070432.1070590,PhysRevResearch.6.043119}, network analysis~\cite{chawla2020discrete,PhysRevResearch.6.043119,paparo2013quantum}, and quantum machine learning~\cite{dernbach2019quantum, ye2023vqne, sato2024qwalkvec}.

Research on the circuit implementation of quantum walks has made progress in recent years~\cite{singh2021quantum,bhavsar2024coined,huerta2020quantum, acasiete2020implementation,wing2023circuit,wing2025circuit,razzoli2024efficient, loke2012efficient,loke2017efficient, nzongani2023quantum, portugal2025efficient, chakraborty2025continuous, PhysRevA.110.052215}.  In the domain of the CTQWs, implementations have traditionally focused on regular structures such as fully connected networks or hypercube lattices.  However, recent works have begun to extend these frameworks to dynamics on complex networks~\cite{portugal2025efficient, chakraborty2025continuous}.  Regarding the DTQWs, two primary approaches have been extensively studied.  The first is Szegedy’s walk~\cite{Szegedy_walk}, which encodes quantum dynamics utilizing transition matrices derived from Markov chains.  This model is inherently adaptable to general networks and has been applied to various network algorithms~\cite{paparo2013quantum,casares2022qfold,ye2023vqne}.  In contrast, the second approach is the coined quantum walk~\cite{10.5555/1070432.1070590}.  The coined quantum walks have mainly focused on regular network structures~\cite{acasiete2020implementation, loke2012efficient, wing2023circuit}.  Although recent studies have explored coined walks on network structures~\cite{singh2021quantum, bhavsar2024coined}, these implementations are typically limited to specific topologies, such as star networks or regular ring lattices.  Consequently, establishing a systematic and resource-efficient circuit architecture applicable to arbitrary complex networks remains an open challenge.

One of the major challenges in building quantum circuits for coined quantum walks on complex networks arises from their irregular structures, which require different coin and shift operators for each node. To address this issue, we previously proposed a circuit implementation that encodes the number of nodes and edges~\cite{10821130}.  The required number of qubits is given by $\lceil \log_2{N} \rceil + \lceil \log_2{|E|} \rceil$. Moreover, the shift operator requires up to $\lceil \log_2{|E|} \rceil$ multi-controlled $X$ (MCX) gates.  This design poses a substantial resource overhead, particularly for dense or scale-free networks where the number of edges $|E|$ significantly exceeds the number of nodes $N$, as seen in Watts–Strogatz~\cite{watts1998collective}, scale-free~\cite{barabasi1999emergence}, and Erd\H{o}s–R\'enyi~\cite{1360016870446635520} network models.

In this study, we propose a resource-efficient circuit design inspired by the construction techniques of Szegedy’s quantum walk to overcome these limitations. The coined walk in our approach is encoded using a dual-register quantum state representation, enabling a simplified shift operator based on SWAP gates. We implement the proposed architecture using Qmod~\cite{vax2025qmod}, a high-level quantum programming language.

Our contribution lies in the development of a scalable and programmable circuit architecture suitable for practical deployment on quantum computing platforms.  We evaluate our framework across multiple classes of complex networks, including Erd\H{o}s–R\'enyi random networks~\cite{1360016870446635520}, small-world networks~\cite{watts1998collective}, and scale-free networks~\cite{barabasi1999emergence}.  Furthermore, to demonstrate the feasibility of our approach on real hardware, we executed the synthesized circuits on the \texttt{ibm\_torino} superconducting quantum processor.  We observed that applying hardware-aware compilation strategies resulted in slight improvements in circuit performance, suggesting that topology-aware circuit design will become increasingly important for implementing graph algorithms on near-term devices.

This paper is organized as follows.  Section~\ref{sec:complex_network} briefly introduces three well-known complex network models used in the experiments.  Section~\ref{sec:preliminaries} formulates the coined quantum walk model on complex networks.  Section~\ref{sec:proposed method} describes the circuit design and implementation.  Section~\ref{sec:result} presents the experimental results.  Section~\ref{sec:discussion} discusses the broader implications of our findings.  Finally, Section~\ref{sec:conclusions} concludes the paper.

\section{Complex Networks}
\label{sec:complex_network}
%about erdos, small world, scale-free
We evaluate the proposed coined quantum walk circuit using three well-known complex network models, the Erd\H{o}s–R\'enyi model~\cite{1360016870446635520}, the Watts–Strogatz model~\cite{watts1998collective}, and the Barab\'asi–Albert model~\cite{barabasi1999emergence}.  Each of these models captures different structural characteristics found in real-world systems.  We describe the generation procedure and define the relevant parameters for each model.

%The Erd\H{o}s–R\'enyi (ER) model $G(N,p)$ generates a random graph with $N$ nodes, where each possible pair of nodes is connected independently with probability $p$~\cite{1360016870446635520}.  This model yields networks with a binomial degree distribution and no strong community or hierarchical structure. Despite its simplicity, the ER model is often used as a baseline for evaluating the behavior of network algorithms under random connectivity.

The Erd\H{o}s–R\'enyi random network $G(N,p)$ is defined on $N$ nodes, where each possible edge between distinct node pairs is independently included with probability $p$~\cite{1360016870446635520}.  A network containing $M$ edges occurs with probability $p^M(1-p)^{\binom{N}{2}-M}$.  The ER network demonstrates that certain structural properties, such as connectivity, emerge sharply at critical probabilities—for instance, a network becomes connected almost surely when $p > \log{N}/N$.

The Watts–Strogatz (WS) model $G(N, k,\beta)$ generates small-world networks characterized by a high clustering coefficient and a short average path length that scales as $L\sim\log N$~\cite{watts1998collective}. The model starts from a ring lattice in which each node is connected to its $k$ nearest neighbors, and then each edge is rewired with probability $\beta$.  As $\beta$ increases from $0$ to $1$, the network transitions from a regular lattice to a random network, retaining high clustering while achieving logarithmically short path lengths.

%The Barab\'asi–Albert (BA) model $G(N,m)$ generates scale-free networks through a growth process governed by preferential attachment~\cite{barabasi1999emergence}.  Starting from a small connected seed network, new nodes are added sequentially, and each new node connects to $m$ existing nodes with a probability proportional to their current degree.  As a result, the model produces networks whose degree distribution follows a power law, $P(k) \sim k^{-\alpha},$ where the exponent typically satisfies ($\alpha \approx 3$). 

The Barab\'asi--Albert (BA) model $G(N,m)$ is a preferential-attachment network model that can generate network structures depending on the parameter $m$~\cite{barabasi1999emergence}.  Starting from a small connected seed network, new nodes are added sequentially, and each new node connects to $m$ existing nodes with probability proportional to their current degree.  The BA model is known to exhibit a power-law degree distribution with an exponent typically close to $\alpha \approx 3$.  In this work, we consider a broad range of $m$ values, namely $m \in [5, N-5]$, in order to evaluate the proposed circuit on diverse irregular network structures rather than restricting attention only to the standard sparse scale-free regime.  Consequently, not all generated network $G(N,m)$ exhibit strict scale-free behavior or a clear power-law degree distribution.

We describe the generation procedure and define the relevant parameters for each model using \texttt{NetworkX}~\cite{SciPyProceedings_11}.  Specifically, we generate three network models using the \texttt{erdos\_renyi\_graph}, \texttt{connected\_watts\_strogatz\_graph}, and \texttt{barabasi\_albert\_graph} functions provided by \texttt{NetworkX}, while varying the corresponding parameters for each model.

\section{Mathematical Model}
\label{sec:preliminaries}
\begin{figure}[t]
    \centering
    \includegraphics[width=0.8\linewidth]{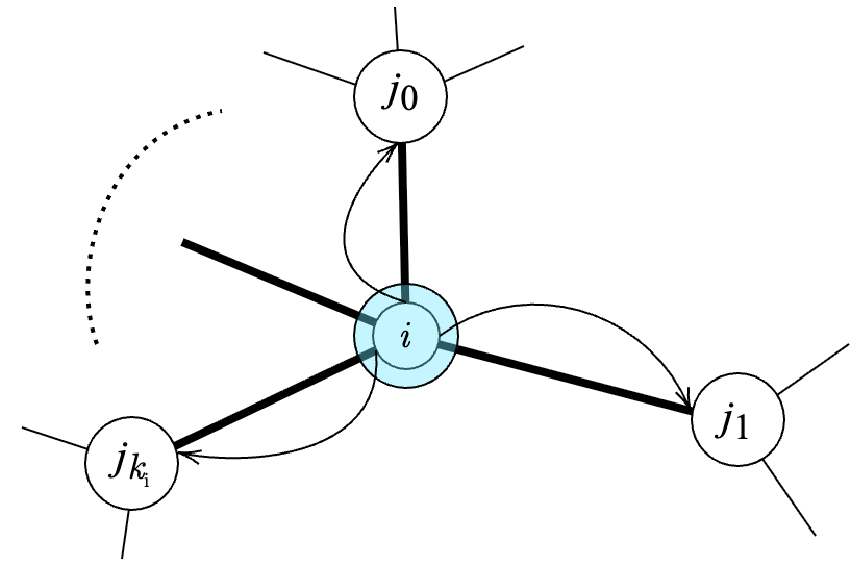}
    \caption{The definition of coined quantum walk on complex networks. $j$ is the neighbor of node $i$. }
    \label{fig:qw_complex}
\end{figure}

We describe a mathematical framework for coined discrete-time quantum walks on general undirected networks, focusing on the definition of the Hilbert space, the coin operator, and the shift operator. 

The complex network is represented as a graph $G = (V, E)$, where $V =\{ 0, 1, 2, \ldots, N-1\}$ is the set of $N$ nodes and $E = \{e_{ij}\}$ is the set of edges.  Fig.~\ref{fig:qw_complex} shows an example of the quantum state of the quantum walker on node $i$.  The quantum state of the walker at time step $t$ is described as
\begin{equation}
    \ket{\psi(t)} = \sum_{i \in V} \sum_{j \in \mathcal{N}(i)} \psi_{ij}(t) \ket{i}\otimes\ket{i \rightarrow j},
    \label{eq:quantum_state}
\end{equation}
where $\mathcal{N}(i)$ denotes the set of neighbors of node $i$, $\psi_{ij}(t)$ denotes the probability amplitude of the walker moving from node $i$ to node $j$ and the basis state $\ket{i \rightarrow j}$ also represents the direction from node $i$ to node $j$.  $k_i$ is the number of edges attached to node $i$.  $\ket{\psi(t)}$ is defined in the Hilbert space $\bigoplus_{i=0}^{N-1}\mathcal{H}_i\otimes\mathcal{H}_{k_i}$, where $\ket{i}\in\mathcal{H}_i$ is associated with the positional degree of freedom, and $\ket{i \rightarrow j} \in H_{k_i}$ is associated with the internal degree of freedom.  $\psi_{ij}(t)$ is the probability amplitude of $\ket{i}\ket{i \rightarrow j}$.

The time evolution of the quantum walker is governed by alternating applications of the coin operator $\hat{C}$ and the shift operator $ \hat{S} $, as
\begin{equation}
    \ket{\psi(t)} = [\hat{S} \hat{C}]^t \ket{\psi(0)}.
    \label{eq:Unitary_evolv}
\end{equation}
The initial state $\ket{\psi(0)}$ is chosen as
\begin{equation}
    \ket{\psi(0)} = \frac{1}{\sqrt{N}} \sum_{i \in V} \frac{1}{\sqrt{k_i}} \sum_{j \in \mathcal{N}(i)} \ket{i} \otimes \ket{i \rightarrow j}.
    \label{eq:initial_state}
\end{equation}
The coined quantum walk can, in general, be initialized from an arbitrary state~\cite{10.5555/1070432.1070590,mukai2020discrete}.  In this work, we choose Eq.~\eqref{eq:initial_state} as a representative initial state for our analysis.

The coin operator $ \hat{C} $ is position-dependent, as the degree $k_i$ may vary for each node $i$.  It is defined as $\hat{C} = \sum_{i} \ket{i}\bra{i}\otimes\hat{C}_i$, where the coin operator $\hat{C}_i$ at node $i$ is given by
\begin{equation}
\hat{C}_i = 2 \ket{s_i} \bra{s_i} - \hat{I}_i,
\label{eq:coin_opeartor}
\end{equation}
where $\ket{s_i} = 1/\sqrt{k_i} \sum_{j \in \mathcal{N}(i)} \ket{i \rightarrow j}$, and $\hat{I}_i$ is the identity operator.  

The shift operator $\hat{S}$ moves the walker from the current node to its neighboring node and updates the internal coin state to indicate the new direction.  It is defined as
\begin{equation}
    \hat{S} \ket{i}\ket{i \rightarrow j} = \ket{j}\ket{j \rightarrow i}.
\end{equation}
This type of shift operation, known as flip-flop shift, is particularly useful for handling networks with varying node degrees, as it naturally adapts to the network's structure.  

Finally, the probability of finding the walker at node $i$ at time step $t$ is given by
\begin{equation}
    P_i(t) = \sum_{j=0}^{k_i-1} \left| \left(\bra{i}\otimes\bra{i \rightarrow j} \right) \ket{\psi(t)} \right|^2.
\end{equation}

\section{Circuit Design}
\label{sec:proposed method}
\begin{figure*}[t]
    \centering
    \includegraphics[width=1.0\linewidth]{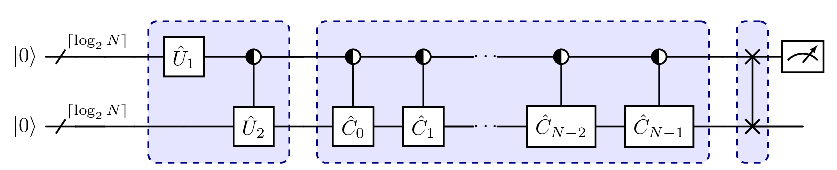}
    \caption{The quantum circuit for implementing the coined quantum walk on complex networks for the walker step as $t=1$.  The first dashed box from the left side represents the initial state of the quantum walker $\ket{\psi(0)}$.  The second dashed-box represents coin operator $\hat{C} = \sum_i\ket{i}\bra{i}\otimes \hat{C}_i$. The third dashed box represents the swap operator as the shift operator $\hat{S}$.  The half-filled circle denotes a control qubit based on its $0/1$ binary state.}
    \label{fig:circuit_DTQW}
\end{figure*}

We propose a systematic method for building the quantum circuits to implement the quantum walks defined in Section~\ref{sec:preliminaries}.  We first encode the quantum state $\ket{i}\otimes\ket{i \rightarrow j}$ using two quantum registers, both representing node labels.
\begin{equation}
    \ket{i}\otimes\ket{i \rightarrow j}\rightarrow
    \ket{i}\ket{j}
\end{equation}

The first and second registers encode the position degree of freedom and the internal degree of freedom of the quantum walker, respectively. The position state is defined as
\begin{equation}
    \hat{U}_1\ket{0}^{n} = \frac{1}{\sqrt{N}}\sum_{i=0}^{N-1}\ket{i},
    \label{eq:gateU1}
\end{equation}
where $i\in\{0,1,\dots,N-1\}$ labels the nodes of the network with $N$ nodes. $n=\lceil\log_2 N\rceil$ is the required number of qubits for encoding nodes.

The internal degree state is defined as
\begin{equation}
    \hat{U}_2^{(i)}\ket{0}^{n} = \frac{1}{\sqrt{k_i}}\sum_{j=0}^{N-1}A_{i,j}\ket{j},
    \label{eq:gateU2}
\end{equation}
where $j\in\{0,1,\dots,N-1\}$ labels the computational basis states of the second register, and $k_i$ is the degree of node $i$. Here, $A_{i,j}$ denotes the $(i,j)$ element of the adjacency matrix of the network, with $A_{i,j}=1$ if $(i,j)\in E$ and $A_{i,j}=0$ otherwise.

Each register is implemented with $n=\lceil \log_2 N \rceil$ qubits and therefore spans a $2^n$-dimensional computational basis. When $N$ is not a power of two, the basis states labeled by $\{N, N+1, \dots, 2^n-1\}$ are not assigned to network nodes. Accordingly, both the position states and the node-dependent internal states are embedded into the $2^n$-dimensional Hilbert space by padding the remaining $2^n-N$ components with zeros. In our construction, the initial state preparation and the subsequent unitary operations are defined so that the quantum walk dynamics remain within the valid subspace spanned by the node labels $\{0,1,\dots,N-1\}$.

We define the controlled operation $C^n(\hat{U}_2)$ by $C^n(\hat{U}_2)\ket{i}\ket{0}^n = \ket{i}\hat{U}_2^{(i)}\ket{0}^n$, where $\hat{U}_2^{(i)}$ acts on the second register conditioned on the control state $\ket{i}$.  The initial state of the quantum walk on the complex network is then given by
\begin{eqnarray}
     \ket{\psi(0)} &=&C^n(\hat{U}_2)(\hat{U}_1\otimes\hat{I})\ket{0}^{n}\otimes\ket{0}^{n}\\
     &=&\frac{1}{\sqrt{N}} \sum_{i=0}^{N-1} \ket{i} \otimes \frac{1}{\sqrt{k_i}} \sum_{j=0}^{N-1} A_{i,j}\ket{j}.
     \label{eq:gate_initial_state}
\end{eqnarray}

The coin operator $\hat{C}$ consists of controlled operation $C^n(\hat{C}_i)$.  The coin operator $\hat{C}_i$ based on the Eq.~(\ref{eq:coin_opeartor}) is given by
\begin{equation}
\hat{C}_i = 2\hat{U}_2\ket{0}^n\bra{0}^n\hat{U}_2^{\dagger}-\hat{I}^n.
\label{eq:gate_Ci}
\end{equation}

%Shift operator
The shift operator $\hat{S}$ shifts the direction of the quantum walk by exchanging the two quantum registers.  Since the quantum state is represented as $\ket{i} \otimes \ket{j}$, the shift operation is defined as
\begin{equation}
\hat{S} \ket{i} \otimes \ket{j} = \ket{j} \otimes \ket{i}.
\label{eq:gate_S}
\end{equation}
This operation can be implemented by swapping each qubit of the first register with the corresponding qubit of the second register using SWAP gates, realizing a flip-flop shift between adjacent nodes.  The full register exchange can be realized by applying an $n$-qubit multi-SWAP operation that swaps all corresponding qubit pairs between the two registers.

%whole circuit
As described in Eq.~(\ref{eq:Unitary_evolv}), one step consists of sequentially applying the coin operator $\hat{C}$ followed by the shift operator $\hat{S}$.  The circuit is composed of two main blocks: the coin operation blocks given and the shift operation block.  By iteratively applying the unitary operation defined in Eq.~(\ref{eq:Unitary_evolv}), multiple steps of the quantum walk can be performed in a straightforward manner.  Fig.~\ref{fig:circuit_DTQW} illustrates the proposed quantum circuit for a single step of the coined quantum walk based on Eqs~(\ref{eq:gateU1})-(\ref{eq:gate_S}).  

To create the quantum circuit, we use Qmod~\cite{vax2025qmod}, a high-level language, and Synthesis~\cite{goldfriend2024design}, which automatically generates the circuit by allocating and optimizing available resources such as the number of qubits and the circuit depth.  Synthesis also allows parameter tuning according to the desired circuit generation and optimization strategy.  For our simple circuit evaluation, we set the parameters as \texttt{transpilation\_option = AUTO\_OPTIMIZE}, \texttt{debug\_mode = False}, and \texttt{optimization\_level = LIGHT}.  See the supplemental material for the Qmod implementation.

\section{Result}
%classiqコンパイルのパラメータ設定を記述する. 
\label{sec:result}

\begin{figure}[t]
    \centering
    \includegraphics[width=0.95\linewidth]{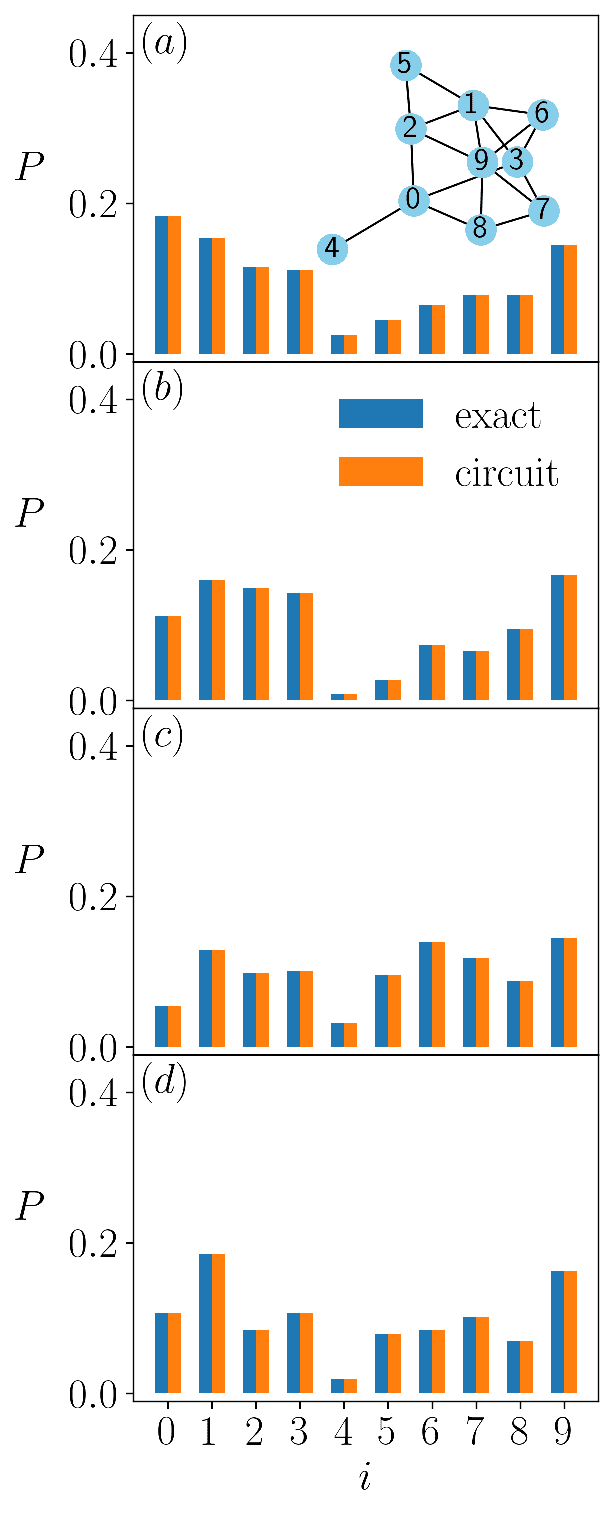}
    \caption{Probability distributions of the quantum walk on an Erd\H{o}s–R\'enyi random network ($n=10$, $p=0.3$). "Exact" and "Circuit" in the legend denote the analytical values and the results obtained from the statevector simulator, respectively. The $y$-axis is shared across all panels. (a) $t=1$, (b) $t=2$, (c) $t=3$, and (d) $t=4$.}
    \label{fig:investigation_toy_problem}
\end{figure}

%以前の手法と比較して改善したものも乗っける
\begin{figure}[t]
    \centering
    \includegraphics[width=0.8\linewidth]{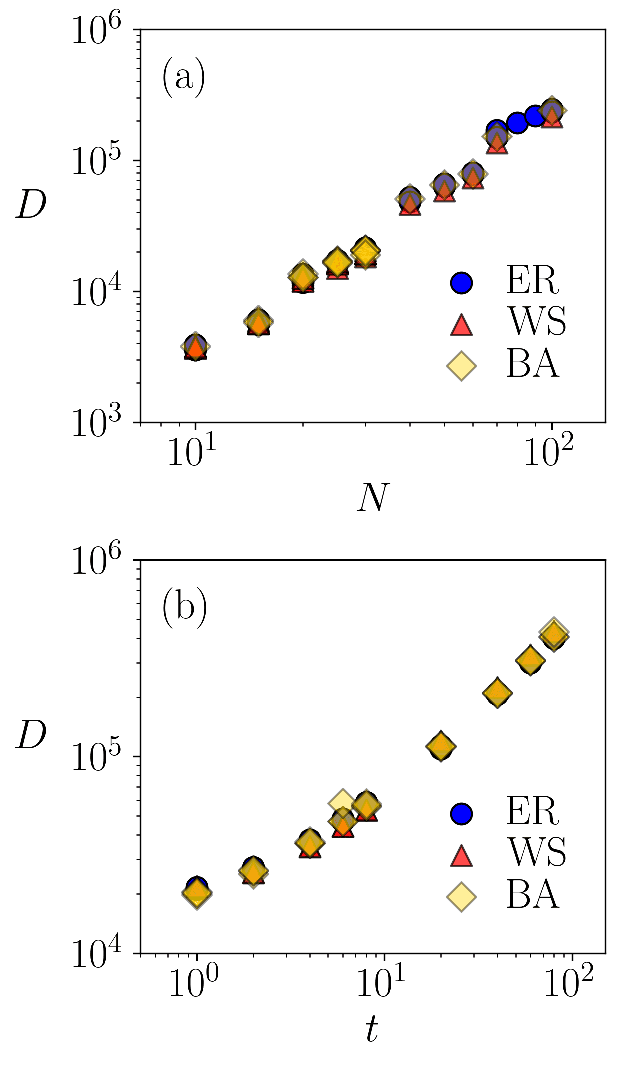}
    \caption{(a) The node-dependent depths of coined quantum walk circuits for the ER model, WS model, and BA model.  The circuits are evaluated at time step $t=1$. (b) The time-dependent depths of coined quantum walk circuits for the ER model, WS model, and BA model.  The circuits are evaluated for $N=2^5$.}
    \label{fig:result_circuit_depth_N_time}
\end{figure}

\begin{figure*}[t]
    \centering
    \includegraphics[width=1.0\linewidth]{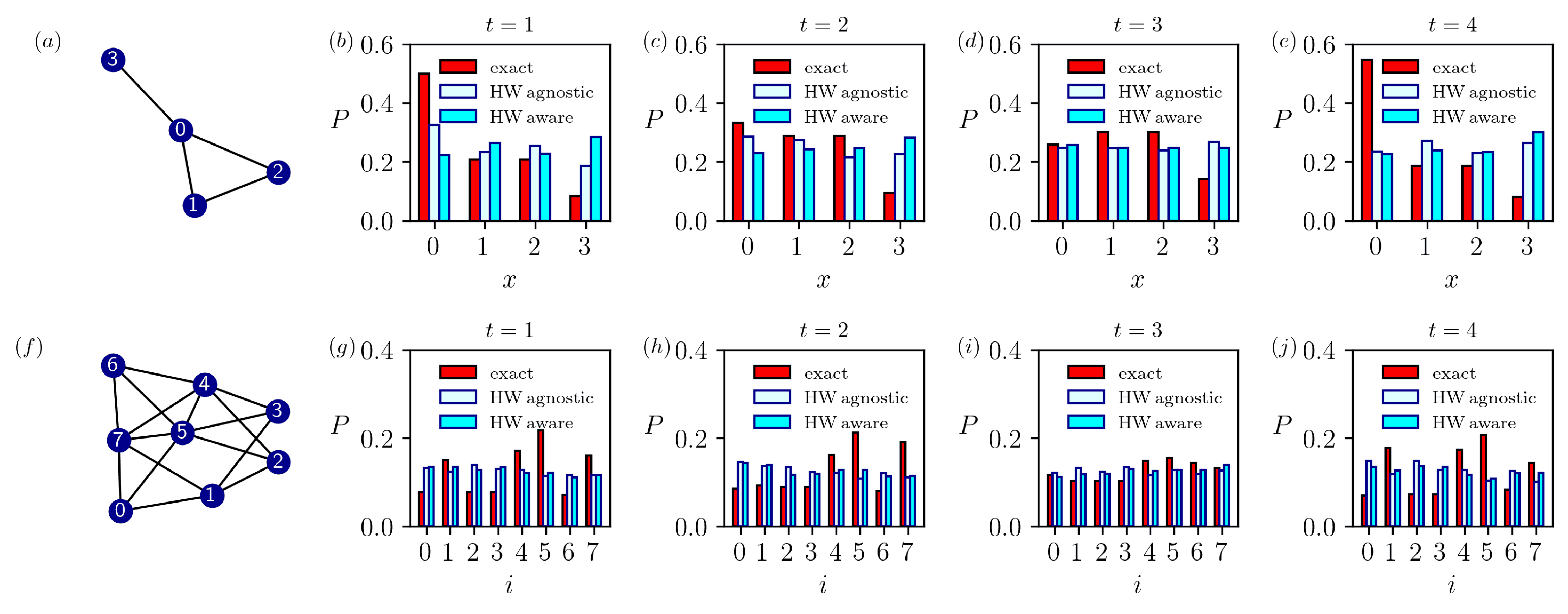}
    \caption{Time-evolution of the probability distribution for a quantum walk on two different WS models.  The histograms compare the exact theoretical simulation (red bars) with experimental results obtained from the \texttt{ibm\_torino} backend (10000 shots).  The experimental data contrast hardware (HW) agnostic (dark blue outline) and HW-aware (cyan filled) synthesis strategies. (a) A WS model with $N=4$, $k=2$, and $\beta=0.2$.  (b)–(e) Probability distributions at time steps $t=1$ to $t=4$ for the network in (a). (f) A WS model with $N=8$, $k=2$, and $\beta=0.2$. (g)–(j) Corresponding probability distributions for time steps $t=1$ to $t=4$.}
    \label{fig:result_hw}
\end{figure*}
% \begin{table}[t]
%     \centering
%     \renewcommand{\arraystretch}{1.2}
%     \begin{tabular}{lcccccc}\hline\hline
%          & $N$ & $|E|$ & $t=1$ & $t=2$ & $t=3$ & $t=4$\\\hline
%         $L_{1}^{\rm agn}$ &4 & 4 & $0.1742..$&$0.1326..$ & $0.1237..$& $0.3126..$\\
%         $L_{1}^{\rm hw}$ &4& 4 &$0.2772..$ & $0.1886..$ & $0.1065..$&$0.3209..$\\
%         Depth$^{\rm agn}$ & 4& 4&$820$&$1079$&$1408$&$1640$ \\
%         Depth$^{\rm hw}$ & 4& 4&$2397$&$3483$&$4354$&$5478$\\\hline
%         $L_{1}^{\rm agn}$ &8 & 16 &$0.2160..$ & $0.2239..$ & $0.08750..$&$0.2517..$\\
%         $L_{1}^{\rm hw}$ &8& 16 &$0.2050..$ & $0.1949..$ & $0.06796..$&$0.2280..$\\
%         Depth$^{\rm agn}$ & 8& 16&$1427$&$1836$&$2357$&$2845$ \\
%         Depth$^{\rm hw}$ & 8& 16&$1396$&$1782$&$2344$&$8186$\\\hline\hline
%         % $L_{1}^{\rm agn}$ &12 & 24 &$ --$ & $ --$ & $--$\\
%         % $L_{1}^{\rm hw}$ &12& 24 &$--$ & $--$ & $--$\\
%         % Depth$^{\rm agn}$ & $12$& $24$& $4406$& $5593$&$7417$&$9121$ \\
%         % Depth$^{\rm hw}$ &12& 24&$16013$& $21394$ & $26832$& $32655$\\\hline\hline
%     \end{tabular}
%     \caption{Experimental results on \texttt{ibm\_torino}. We compare the total variation distance, $L_1$, and circuit depth between hardware-agnostic (agn) and hardware-aware (hw) synthesis approaches for different time steps $t$.}
%     \label{tab:L1_results}
% \end{table}

We first evaluated the proposed circuit for ER, WS, and BA models.  For each network type, we generated networks with node sizes $N\in[10,100]$.  We set parameters $p$ as $p = [0.2$, $0.4$, $0.6$, $0.8$, $1.0$], $\beta\in[0,1]$, $k=4$, and $m\in[5,N-5]$ with increments of $5$ for the ER model, WS model, and BA model, respectively.  For each network data, we generate three random instances and evaluate the average and standard deviation of the results to account for network randomness.  

We investigate the correctness of the proposed circuit by comparing the theoretical value for the small networks.  Fig.~\ref{fig:investigation_toy_problem} shows the probability distribution of the quantum walk on the ER model with $N=10$ and $p=0.3$.  As a benchmark, the theoretical value is calculated using classical matrix-based simulation of the quantum walk dynamics. We confirm that the quantum circuit simulation results matched well with the theoretical values.

Fig.~\ref{fig:result_circuit_depth_N_time}(a) illustrates the scaling behavior of the proposed circuit depth with respect to the number of nodes in three complex network models.  Across all three networks, the circuit depth consistently exhibited around $N^\alpha$ ($\alpha=1.9$) growth as the number of nodes increased, given by 
\begin{eqnarray}
    D_{\rm ER} &=& 38(12)N^{1.91( 7)}\\
    D_{\rm WS} &=& 41(8)N^{1.86(4)}\\
    D_{\rm BA} &=& 38(12)N^{1.90(7)}
\end{eqnarray}
for ER, WS and BA models, respectively.  

Fig.~\ref{fig:result_circuit_depth_N_time}(b) illustrates the scaling behavior of the proposed circuit depth with respect to the number of steps in three complex network models.  Across all three network, the circuit depth given by 
\begin{eqnarray}
    D_{\rm ER} &=& 8969(669)t^{0.86(1)}\\
    D_{\rm WS} &=& 8841(743)t^{0.88(2)}\\
    D_{\rm BA} &=& 8841(743)t^{0.88(2)}
\end{eqnarray}
for ER, WS and BA models, respectively.  

This trend remained stable regardless of the specific structural parameters of the networks, such as the connection probability $p$, the rewiring probability $\beta$, or the attachment parameter $m$.  The observed consistency across different models suggests that the proposed quantum circuit design is robust to topological variations in complex networks.  This linear depth scaling is particularly important for fault-tolerant quantum computing, as circuit depth directly affects the feasibility of implementation on current quantum hardware. These results indicate that the proposed framework can reliably support a broad range of complex network structures without incurring significant additional complexity, thereby offering a scalable and generalizable approach to implementing coined quantum walks in quantum circuits.

\begin{table}[t]
    \centering
    \caption{Experimental results on \texttt{ibm\_torino}. We compare the total variation distance $L_1$, Hellinger fidelity $F_{\rm H}$, and circuit depth between hardware-agnostic (agn) and hardware-aware (hw) synthesis approaches for different time steps $t$.}
    \renewcommand{\arraystretch}{1.0}
    \scalebox{0.93}{
    \begin{tabular}{ccccccccc}\hline\hline
        $N$ & $|E|$ & $t$ 
        & $L_{1}^{\rm agn}$ & $L_{1}^{\rm hw}$ 
        & $F_{\rm H}^{\rm agn}$ & $F_{\rm H}^{\rm hw}$ 
        & Depth$^{\rm agn}$ & Depth$^{\rm hw}$ \\\hline
        \multirow{4}{*}{4} & \multirow{4}{*}{4}
        & 1 & $0.1742$ & $0.2772$ & $0.9587$ & $0.8849$ & $820$  & $2397$ \\
        & & 2 & $0.1326$ & $0.1886$ & $0.9642$ & $0.9370$ & $1079$ & $3483$ \\
        & & 3 & $0.1237$ & $0.1065$ & $0.9740$ & $0.9803$ & $1408$ & $4354$ \\
        & & 4 & $0.3126$ & $0.3209$ & $0.8757$ & $0.8594$ & $1640$ & $5478$ \\\hline
        \multirow{4}{*}{8} & \multirow{4}{*}{16}
        & 1 & $0.2160$ & $0.2050$ & $0.9474$ & $0.9519$ & $1427$ & $1396$ \\
        & & 2 & $0.2239$ & $0.1949$ & $0.9460$ & $0.9590$ & $1836$ & $1782$ \\
        & & 3 & $0.08750$ & $0.06796$ & $0.9907$ & $0.9944$ & $2357$ & $2344$ \\
        & & 4 & $0.2517$ & $0.2280$ & $0.9302$ & $0.9409$ & $2845$ & $8186$ \\\hline\hline
    \end{tabular}
    }
    \label{tab:L1_Hl_results}
\end{table}

%ION-Q
%IBM_brisbane/heroin ? fidelityで 
%hw agnosticとwithout agnotsticで評価
Next, we validate the feasibility of our approach on actual quantum hardware. We executed the synthesized circuits on \texttt{ibm\_torino}, a $133$-qubit superconducting processor. To quantify the agreement between the hardware results and the exact simulation, we evaluate the output probability distributions using both the total variation distance, $L_1$~\cite{venegas2012quantum}, and the Hellinger fidelity~\cite{PhysRevA.97.062342}, $F_{\rm H}$, as complementary benchmarks. The total variation distance is defined as
\begin{equation}
L_1(t) = \frac{1}{2}\sum_x \bigl| P_{\rm hw}(t,x) - P_{\rm exact}(t,x) \bigr|,
\end{equation}
and the Hellinger fidelity is defined as
\begin{equation}
F_{\rm H}(t) = \left( \sum_x \sqrt{P_{\rm hw}(t,x)\,P_{\rm exact}(t,x)} \right)^2,
\end{equation}
where $P_{\rm hw}(t,x)$ and $P_{\rm exact}(t,x)$ denote the probability distributions obtained from the real hardware execution and the exact state-vector simulation, respectively. %Smaller values of $L_1$ and larger values of $F_{\rm H}$ indicate better agreement between the two distributions.

To investigate the quantum circuits used in these experiments, we employed hardware-aware synthesis to optimize the circuits under the constraints of the \texttt{ibm\_torino} processor and compared the resulting circuits against a hardware-agnostic baseline.  We conducted experiments on WS models with $N=4$ and $N=8$, and Fig.~\ref{fig:result_hw} shows the probability distributions of the quantum walk at time steps $t=1,2,3$, and $4$.  Although the experimental distributions show quantitative deviations from the exact simulation due to noise in current NISQ devices, clear differences can be observed between the hardware-agnostic and hardware-aware synthesis strategies, as summarized in Table~\ref{tab:L1_Hl_results}.

For the WS model with $N=4$, the hardware-aware synthesis results in deeper circuits and worse agreement with the exact distribution than the agnostic approach, as reflected by larger $L_1$ values and smaller $F_{\rm H}$ values for most time steps.  This larger circuit depth is mainly due to additional routing operations, such as SWAP gates, required by the connectivity and qubit allocation constraints of \texttt{ibm\_torino}.  This suggests that, for such a small network, the overhead introduced by hardware constraints can outweigh the benefit of hardware-aware optimization.

In contrast, for the WS model with $N=8$, the hardware-aware synthesis shows better agreement with the exact distribution across all tested time steps.  As shown in Table~\ref{tab:L1_Hl_results}, the hardware-aware circuits consistently achieve slightly smaller $L_1$ values and slightly larger $F_{\rm H}$ values than the agnostic baseline.  Although the hardware-aware circuits also have larger circuit depths, this is likely because the synthesis explicitly accounts for the device topology and qubit usage constraints of \texttt{ibm\_torino}, which can increase the number of routing gates. These results suggest that hardware-aware synthesis can improve the performance of quantum walk circuits on current devices, even when such improvements are not directly reflected by a simple circuit-depth comparison.

\section{Discussion}
\label{sec:discussion}
\begin{table}[t]
    \centering
    \caption{Comparison of resource requirements.  The table lists the depth and width of the circuit and the gate complexity for the coin and shift operators.}
    \renewcommand{\arraystretch}{1.2}
    \begin{tabular}{lcc}\hline\hline
        & Previous work~\cite{10821130} & Proposed method \\ \hline
        Circuit width & $\lceil\log_2{N}\rceil+\lceil\log_2{|E|}\rceil$ & $2\lceil\log_2{N}\rceil$ \\
        Coin operator & $\mathcal{O}(N)$ & $\mathcal{O}(N)$ \\
        Shift operator & $\mathcal{O}(|E| \cdot \text{poly}(\log N))$ & $\mathcal{O}(\log N)$ \\ \hline\hline
    \end{tabular}
    \label{tab:comparison_circuit}
\end{table}
The experimental results demonstrate that the circuit depth exhibits consistent scaling behavior with respect to the number of nodes for the three network models.  As shown in Fig.~\ref{fig:result_circuit_depth_N_time}, the circuit depth scales as approximately $N^{1.9}$ for ER, WS, and BA models.  This result suggests that the proposed method is robust to topological variations in complex networks.  The scaling exponent of $\alpha \approx 1.9$ indicates that the circuit complexity is mainly determined by the state preparation $\hat{U}_2$ or node-dependent coin operators $\hat{C}_i$.  The quantum walk on arbitrary complex networks requires the polynomial gate complexity.

In the Qmod implementation, the behavior of scaling law, $N^{1.9}$, can be understood from the construction of the position-dependent coin operator.  While the flip-flop shift $\hat{S}$ is implemented by swap gates and therefore has depth only $\mathcal{O}(\log N)$, the coin operator requires, for each node $i$, a reflection about the superposition defined in Eq.~\eqref{eq:coin_opeartor}.  Accordingly, the coin operator is implemented by iterating over all nodes and sequentially applying the corresponding controlled state-preparation blocks. Its depth therefore scales as
$D(N) \sim \sum_{i=0}^{N-1} D_{\mathrm{prep}}(i) + \mathcal{O}(N\log N),$ where the second term accounts for the swap gates.  Furthermore, if $|s_i\rangle$ is prepared using a general state preparation on $n=\lceil \log_2 N\rceil$ qubits, then $D_{\mathrm{prep}}(i)=\mathcal{O}(2^n)=\mathcal{O}(N),$ since $2^n$ is proportional to $N$ up to a constant factor.  This yields an overall depth scaling of $D(N)=\mathcal{O}(N^2),$ consistent with generic state-preparation bounds~\cite{plesch2011quantum} and broadly consistent with the empirically observed scaling of approximately $N^{1.9}$.

For the scaling with respect to the number of walk steps $t$ for a fixed $N$, the approximately linear behavior is governed by $D(t)\approx D_0 + t D_{\rm walk}$ where $D_0$ is the depth of state preparation and $D_{\rm walk}$ is quantum walk operator, since the same step operator is repeated $t$ times.  However, the depth compiled after Classiq synthesis includes compilation-dependent optimizations, so fitting the data $\mathcal{O}(t^{0.88})$ in a finite range of $t$ by a power law can produce an effective exponent smaller than one.  

%Resource Comparison with Previous Methods
We also compare the resource requirements of our previous approach~\cite{10821130}.  The previous method required $\lceil\log_2{N}\rceil+\lceil\log_2{|E|}\rceil$ qubits because it encodes both node and edge information.  In contrast, our new approach requires the width of our circuit is given by $2\lceil\log_2{N}\rceil$, as we use two registers proportional to $N$ for encoding node information and coin state.   For the three models in this study, the number of edges $E$ is equal to or greater than the number of nodes $N$ ($|E| \ge N$), and realizes a more efficient resource.  This means the proposed circuit can be implemented using fewer qubits than previous approaches in all cases. 

Second, the proposed method significantly reduces the total number of gates. In earlier methods, the shift operation relied on up to $E$ multi-controlled X (MCX) gates.  Since an MCX gate requires more CNOTs as the number of control qubits increases, the total CNOT count scales approximately as $\mathcal{O}(|E| \cdot \text{poly}(\log N))$.  In contrast, the proposed circuit employs SWAP gates for the shift operation, requiring a count proportional to $\lceil\log_2{N}\rceil$.  Since each of these gates decomposes into three CNOTs, the total complexity becomes $3\times\lceil\log_2{N}\rceil=\mathcal{O}(\log N)$.  This leads to a major reduction in resources, especially for dense complex networks.

%Feasibility on Quantum Hardware
We also validate the feasibility of our approach using \texttt{ibm\_torino}. The experimental results show that the impact of hardware-aware optimization varies depending on the network scale. For the $N=8$ network, the hardware-aware synthesis improves the agreement with the exact distribution, as reflected by smaller $L_1$ values and larger Hellinger fidelity $F_{\rm H}$ values.  In contrast, for the smaller network with $N=4$, the overhead from qubit connectivity constraints becomes significant and outweighs the benefits of optimization. This indicates that while the proposed circuit is scalable in theory, the limited connectivity of current NISQ devices remains a challenge. Therefore, topology-aware circuit design becomes increasingly important as the network size and circuit complexity increase.

%Future Outlook
Finally, we discuss the hardware requirements for practical applications. Based on the scaling law $D \approx 40 N^{1.9}$ derived from our results, implementing the coined quantum walk on a network of size $N=100$ requires a circuit depth of approximately $2.5 \times 10^5$.  This depth exceeds the capacity of current NISQ devices and implies a per-gate error threshold of $\epsilon \sim 10^{-5}$.  This threshold is estimated from the rough approximation that the accumulated error probability scales as $ \epsilon D $ and should remain well below unity for meaningful computation~\cite{Nielsen_Chuang_2010}.  Since state-of-the-art physical qubits typically exhibit error rates around $10^{-3}$ to $10^{-4}$~\cite{google2023suppressing}, executing such deep circuits, $D \sim 10^5$, is infeasible on NISQ devices and requires the use of logical qubits.

\section{Conclusions} 
\label{sec:conclusions}
In this work, we proposed a programmable quantum circuit framework for coined discrete-time quantum walks on complex networks. By introducing a dual-register encoding, we significantly reduced the gate complexity of the shift operator, which had been a bottleneck in previous work, and thereby enabled a more efficient implementation. We evaluated the proposed framework on Erd\H{o}s--R\'enyi, Watts--Strogatz, and Barab\'asi--Albert models, and found that the circuit depth scales approximately as $N^{1.9}$ across these network types. We also executed the circuits on the \texttt{ibm\_torino} superconducting quantum processor to examine the feasibility of the approach on real hardware.  For small networks, connectivity constraints introduced additional overhead, while for larger networks the optimization improved the results.  These findings suggest that topology-aware circuit design becomes more important as the network size increases.  Overall, our framework provides a flexible method for studying quantum dynamics on complex networks.  Although current NISQ devices are still limited to small-scale demonstrations, our scaling results suggest that applications to medium-sized networks may become feasible in the early fault-tolerant era.

\bibliographystyle{unsrt}  
\bibliography{main}
\clearpage
\onecolumngrid
\appendix

\begin{center}
  \textbf{\large Supplemental Material for "Coined Quantum Walks on Complex Networks for Quantum Computers"}
\end{center}
\vspace{1cm}

\section{Implementation with Qmod}
Here, we provide a step-by-step explanation of the Qmod implementation corresponding to the mathematical model of the coined quantum walk described in the main text. The implementation leverages Qmod's high-level abstractions to automatically synthesize circuits for arbitrary complex networks.

%\subsection{Data Structure and Graph Definition}
First, we define the complex network structure using \texttt{networkx}. For this example, we generate a Watts-Strogatz graph. We also prepare helper functions to extract node degrees and neighbor information, which correspond to the adjacency matrix elements $A_{i,j}$ and degree $k_i$ in the theoretical model.

\begin{lstlisting}
from classiq import *
import numpy as np
import networkx as nx

N = 8
k = 2
beta = 0.5
G = nx.watts_strogatz_graph(N, k, beta)
num_qubits = int(np.ceil(np.log2(N)))

def get_edges_of_node(G, i):
    return [j for j in G.neighbors(i)]

def inner_degree(G, i):
    """
    Constructs the probability distribution for the neighbor states of node i.
    This corresponds to the uniform superposition over neighbors in Eq. (3).
    """
    l_array = np.zeros(2**num_qubits)
    neighbors_list = get_edges_of_node(G, i)
    k = len(neighbors_list)
    for j in neighbors_list:
        l_array[j] = 1
    # Normalized probability for state preparation
    return l_array / k
\end{lstlisting}

\subsection{State Preparation}
The preparation of the state involves two steps. The operator $\hat{U}_1$ creates a uniform superposition over all nodes in the first register \texttt{x} (representing node $i$).
\begin{equation}
    \hat{U}_1\ket{0}^{n} = \frac{1}{\sqrt{N}}\sum_{i=0}^{N-1}\ket{i}
\end{equation}
In Qmod, if $N$ is a power of $2$, this is a simple Hadamard transform. Otherwise, we use \texttt{inplace\_prepare\_state} with a uniform probability distribution.

The operator $\hat{U}_2^{(i)}$ prepares the coin state with a superposition of neighbors for a specific node $i$ in the second register \texttt{y} (representing neighbor $j$).
\begin{equation}
    \hat{U}_2^{(i)}\ket{0}^{n} = \frac{1}{\sqrt{k_i}}\sum_{j=0}^{N-1}A_{i,j}\ket{j}
\end{equation}
This is implemented using a \texttt{control} block iterating over all nodes $i$. For each $i$, the target operation \texttt{inplace\_prepare\_state} initializes register \texttt{y} based on the local connectivity defined by \texttt{inner\_degree($G$, $i$)}.

\begin{lstlisting}
@qfunc
def prepare_initial_state(x: QNum[num_qubits], y: QNum[num_qubits]):
    # Implementation of U1 on register x
    if N == 2**num_qubits:
        hadamard_transform(x)
    else:
        prob_array = np.ones(2**num_qubits) / N
        prob_array[N:2**num_qubits] = 0
        inplace_prepare_state(prob_array.tolist(), 0.0, x)
    
    # Implementation of controlled-U2 on register y
    for i in range(N):
        control(
            x == i,
            lambda: inplace_prepare_state(inner_degree(G, i).tolist(), 0.0, y),
        )
\end{lstlisting}

\subsection{Coin Operator}
The coin operator $\hat{C}$ applies a local Grover diffusion operator at each node. Mathematically, for each node $i$, the local coin $\hat{C}_i$ is:
\begin{equation}
    \hat{C}_i = 2\hat{U}_2^{(i)}\ket{0}\bra{0}(\hat{U}_2^{(i)})^{\dagger} - \hat{I}
\end{equation}
In Qmod, the \texttt{grover\_diffuser} function automatically implements the reflection $2\ket{s}\bra{s} - I$ given a state preparation oracle. Here, the oracle is exactly the \texttt{inplace\_prepare\_state} function used for $\hat{U}_2^{(i)}$. We wrap this in a \texttt{control} block to apply the specific diffuser for each node position $i$ (corresponding to register \texttt{x}).

\begin{lstlisting}
@qfunc
def my_coin(x: QNum[num_qubits], y: QNum[num_qubits]):
    for i in range(N):
        control(
            x == i,
            stmt_block=lambda: grover_diffuser(
                lambda y: inplace_prepare_state(inner_degree(G, i).tolist(), 0.0, y),
                y
            )
        )
\end{lstlisting}

\subsection{Shift Operator}
The shift operator performs a flip-flop shift by swapping the position and direction registers.
\begin{equation}
    \hat{S} \ket{i}\ket{j} = \ket{j}\ket{i}
\end{equation}
This corresponds to a swap of the two quantum registers \texttt{x} and \texttt{y}. In Qmod, this is concisely expressed using the \texttt{multiswap} gate.

\begin{lstlisting}
@qfunc
def my_shift(x: QNum[num_qubits], y: QNum[num_qubits]):
    multiswap(x, y)
\end{lstlisting}

\subsection{Whole Circuit}
Finally, the discrete-time quantum walk is constructed by iteratively applying the coin and shift operators. The \texttt{power} function in Qmod efficiently implements the repetition of the step unitary $U^t = (\hat{S}\hat{C})^t$.

\begin{lstlisting}
@qfunc
def discrete_quantum_walk(
    time: CInt,
    coin_qfuncs: QCallable[QNum, QNum],
    shift_qfuncs: QCallable[QNum, QNum],
    x: QNum,
    y: QNum
):
    power(
        time,
        lambda: (
            coin_qfuncs(x, y),
            shift_qfuncs(x, y),
        ),
    )

@qfunc
def main(x: Output[QNum[num_qubits]]):
    y = QNum("y", num_qubits)
    allocate(num_qubits, x)
    allocate(num_qubits, y)
    
    # Initialize state |psi(0)>
    prepare_initial_state(x, y)
    
    # Execute walk for t=1 step
    discrete_quantum_walk(1, my_coin, my_shift, x, y)
\end{lstlisting}

\subsection{Circuit Synthesis}
We employ two distinct synthesis strategies to generate the final quantum circuits: hardware-agnostic synthesis for simulation and hardware-aware synthesis for execution on real devices.

For numerical simulations and logical verification, we generate the circuit without targeting a specific backend. This approach optimizes for general logical depth and qubit count, independent of physical constraints.

\begin{lstlisting}
# Standard Synthesis
qmod = create_model(main)
qprog = synthesize(qmod)
\end{lstlisting}

For the experimental validation on the \texttt{ibm\_torino} processor, we utilize hardware-aware synthesis. By defining a \texttt{Preferences} object with the specific backend provider and name, the synthesis engine optimizes the gate decomposition and routing to match the device's connectivity and basis gate set. This step is crucial for minimizing the error rates discussed in the experimental results.

\begin{lstlisting}
# Hardware-aware Synthesis
preferences = Preferences(
    backend_service_provider="IBM Quantum",
    backend_name="ibm_torino"
)

# Synthesize with hardware constraints
qmod_hw = create_model(main, preferences=preferences)
qprog_hw = synthesize(qmod_hw)
\end{lstlisting}

\subsection{Visualization}
\begin{figure}[tb]
    \centering
    \includegraphics[width=0.8\linewidth]{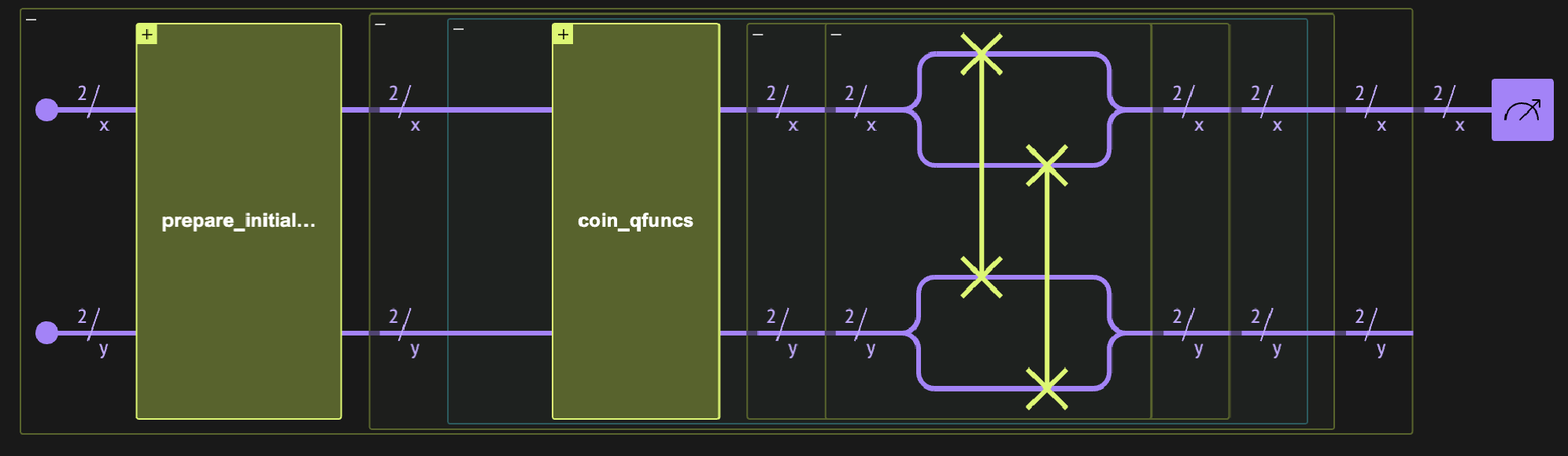}
    \caption{Quantum circuit for the coined quantum walk synthesized using the Classiq Platform. The circuit implements the walk dynamics on a complex network with $N=4$ nodes. The first and second blocks represent the initial state preparation and the coin operator, respectively. The \texttt{swap\_gate} block corresponds to the flip-flop shift operator.}
    \label{fig:vizualize}
\end{figure}
Finally, Qmod provides built-in tools to visualize the synthesized quantum circuit and analyze its properties. The \texttt{show()} function opens an interactive circuit viewer, allowing for a hierarchical inspection of the block-encoded operators, as shown in Fig.~\ref{fig:vizualize}. This command generates the quantum circuit diagrams presented in the main text, confirming that the high-level functional definitions are correctly compiled into the gate-level implementation.

\begin{lstlisting}
# Visualize the generated circuit
show(qprog)

# (Optional) Analyze circuit properties such as depth
# print(qprog.transpiled_circuit.depth)
\end{lstlisting}

\end{document}